\documentclass[final,1p,times]{elsarticle}

\usepackage[T1]{fontenc}
\usepackage{lmodern}
\usepackage{pbox}
\usepackage{array}
\usepackage{makecell}
\usepackage{multirow}
\usepackage{dsfont}
\usepackage{url}
\usepackage{subfig}
\usepackage{amsmath,amsfonts,amssymb, amsthm,bm}
\usepackage{graphicx}
\usepackage{float}
\usepackage{booktabs,tabularx}

\usepackage{stfloats}
\usepackage[toc,page]{appendix}

\usepackage{pbox}
\usepackage{array}

\usepackage{lineno}

\begin{document}

\begin{frontmatter}

\title{Crop yield probability density forecasting via quantile random forest and Epanechnikov Kernel function}

 \author[label1]
 {Samuel Asante Gyamerah}
 \author[label2]
 {Philip Ngare}
 \author[label3]
 {Dennis Ikpe}
 \address[label1]{Pan African University, Institute for Basic Sciences, Technology, and Innovation, Kenya}
\address[label2]{University of Nairobi, Kenya}
\address[label3]{Michigan State University, USA}
 \tnotetext[label1]{correspondence:\,\, saasgyam@gmail.com}

\begin{abstract}	
A reliable and accurate forecasting model for crop yields is of crucial importance for efficient decision-making process in the agricultural sector. However, due to weather extremes and uncertainties, most forecasting models for crop yield are not reliable and accurate. For measuring the uncertainty and obtaining further information of future crop yields, a probability density forecasting model based on quantile random forest and Epanechnikov kernel function (QRF-SJ) is proposed. The nonlinear structure of random forest is applied to change the quantile regression model for building the probabilistic forecasting model. Epanechnikov kernel function and solve-the equation plug-in approach of Sheather and Jones are used in the kernel density estimation. A case study using the annual crop yield of groundnut and millet in Ghana is presented to illustrate the efficiency and robustness of the proposed technique. The values of the prediction interval coverage probability and prediction interval normalized average width for the two crops show that the constructed prediction intervals capture the observed yields with high coverage probability. The probability density curves show that QRF-SJ method has a very high ability to forecast quality prediction intervals with a higher coverage probability. The feature importance gave a score of the importance of each weather variable in building the quantile regression forest model. The farmer and other stakeholders are able to realize the specific weather variable that affect the yield of a selected crop through feature importance. The proposed method and its application on crop yield dataset are the first of its kind in literature. 
\end{abstract}

\begin{keyword}
 climate change \sep crop yield uncertainty \sep crop yield forecasting \sep quantile random forest \sep kernel density estimation \sep Epanechnikov kernel 
\end{keyword}

\end{frontmatter}


\section{Introduction}
\label{S:1}
The agriculture sector is seen as one of the biggest emitter of greenhouse gases and concurrently a major sector that is affected by climate change. In reviewing the factors that affect crop growth, productivity, and yield, \cite{rana2014} indicated that soil moisture, the availability of soil nutrients, and solar radiation are the top three factors that limit the growth of crops and hence limit the yield of crops. Changes in the surface temperature, humidity, and rainfall affects the moisture content of the soil and the level of nutrients in the soil \cite{brevik2013potential}. Hence, there is a direct effect of climate change on crop growth, productivity and yield \citep{chen2004yield,isik2006analysis,southworth2000consequences}. This has significantly affected the yield of most crops causing economic and food security risks in most developing and under-developed countries \citep{asfaw2016managing,gornall2010implications}. 
\\
[1mm]
Small-scale farmers in most developing and under-developed countries cannot mitigate the effects of climate variability \citep{asfaw2016managing}. This mostly have great impact on the farming activities and economy of these farmers. For this reason, an effective and reliable insurance is needed to hedge farmers and stakeholders from the peril of weather uncertainties. Traditional insurance for agricultural risk management is not patronized in most developing countries because of high premiums, loss adjustments, moral hazards, adverse selections, and complex information requirements \citep{gyamerah2018regime}. However, weather derivatives and index-based insurance such as area-yield and weather index insurance are seen as effective risk management tools in the agricultural sector for both small- and large-scale farmers in developing/under-developed countries. Accurate forecasting of crop yields is a principal component for the ratemaking process in the derivative and index-insurance markets. An accurate and mathematically tractable crop yield forecasting model is important for the farmer, policymakers, the government, field managers, and industry players in decision making process \cite{delince2017recent}. 
\\
[2mm]
To improve the  performance of the methods used in crop yield forecasting, a lot of research  have been done in recent decades. A number of literatures based on statistical models have been used to predict the yield of crops  \citep{shi2013review, michler2015risk, lobell2010use}. These literatures have served as a substitute to process-based models, which always involve a comprehensive data on the conditions of the soil, cultivar, and management. \cite{choudhury2014crop} used different time series models (simple and double exponential smoothing, Damped-Trend Linear Exponential Smoothing and autoregressive moving averages (ARMA)) to predict maize yield in five communities in Ghana. The authors concluded that the ARMA model was more robust than the other time-series models. However, time series models such as moving averages, simple and double exponential smoothing, quadratic and linear regression perform poorly in predicting crop yields \citep{hong2008variations, zhang2007application}. These statistical predictions suffer from different sources of error like variations in weather variables. In most statistical methods, there are always little or no interaction between the weather variables used for prediction. However, crop yield and weather variables are highly nonlinear and there are interactions between weather variables \cite{yu2011estimating, chmielewski1995relationship}. Hence, using statistical methods will be computationally costly and may not lead to an optimal performance especially when there are cases of extreme events. Alternative to statistical models, are the emergence of machine learning (ML) techniques such as random forest, support vector regression, and neural networks. These ML techniques are able to capture the nonlinearity of crop yields and its predictors \citep{jeong2016random,everingham2016accurate,dahikar2014agricultural}. Hybrid methods between statistical and machine learning approach are seen to improve the accuracy of predictions in most forecasting problems \citep{fang2018quantile,jiang2017semi}.
\\
[1mm]
Generally, most statistical and machine learning literatures on crop yield forecasting are based on point forecasts \citep{michler2015risk,choudhury2014crop,jeong2016random,everingham2016accurate}. Point forecasts give an estimate of the future crop yield for each time horizon and do not convey any information about the uncertainty of the predictions. Alternative to point forecasting is prediction interval. Prediction interval builds a well calibrated upper and lower bound between which the future unknown prediction value is expected to lie with a certain prescribed probability \cite{chatfield2000time}. Due to the increased uncertainty of weather in recent years as a result of climate change \citep{deser2012uncertainty, reilly2001uncertainty}, point and interval predictions are not able to predict the yield of crops accurately. 
Unlike point prediction and prediction interval, probability density forecasting can measure the uncertainty of crop yields by building probability density function of the forecasting results. With the growing increase in climate change and its effects on crop yields, it is imperative to measure the uncertainties associated with crop yield forecast. 
\\
In probability density forecasting, Kernel density estimation (KDE) is very important in the density estimation process. KDE is a non-parametric method of estimating the distribution of a dataset without prior assumptions of the datasets \cite{wand1994kernel}. Appropriate choice of bandwidth for a kernel density estimator is of crucial importance to the density function of random variables. Quantile regression (QR) can be used to construct a nonparametric probability density forecast. Given one or more covariates, QR generalizes the theory of a univariate quantile to a conditional quantile. Because of the robustness of QR in handling outliers in explained measurements, it is widely used for regression analysis in the areas of econometrics and statistics \citep{barnwal2013climatic}. Conventional linear QR is however unable to deal with complex non-linear problems \citep{wang2012bayesian}. To explore non-linear functions for QR, \citep{meinshausen2006quantile} proposed a quantile random forest (QRF) model, which combines the advantages of random forest and quantile regression models. In furtherance to the application of \citeauthor{meinshausen2006quantile} proposed model, \cite{jiang2017semi} proposed a hybrid semi-parametric quantile regression forest to estimate the non-linear relationship in multi-period value-at-risk. They concluded that their proposed method was more accurate compared to common distributions like normal distribution. In the area of medicine, \cite{fang2018quantile} applied quantile regression forest to Cancer Cell Line Encyclopedia (CCLE) dataset to give a point and interval prediction. The quantile regression forest improved the accuracy of prediction of drug response. However, these literatures \citep[e.g.][]{fang2018quantile} are not able to obtain the probability density functions of the future response variable in a single QRF model. Hence the need to combine quantile regression forest and kernel density estimation to estimate the probability density functions of future crop yields. To obtain a complete crop yield probability density curve, Epanechnikov kernel function and solve-the-equation plug-in approach of Sheather and Jones (SJ) bandwidth selection method \cite{jones1996brief} are combined with quantile random forest (QRF) model. Our proposed method (QRF-SJ) will help in obtaining a complete conditional probability density for different time horizons by selecting a suitable bandwidth and kernel function. 
\\
[1mm]
To ascertain the robustness of QRF for the proposed QRF-SJ model, a comparative analysis between QRF and other ML forecasting techniques (Support vector regression, radial basis neural network, radial basis neural network, generalized linear model) are presented. The contributions of this paper are: 1) We implement a comprehensive probabilistic crop yield forecasting method based on quantile random forest and Kernel density estimation (QRF-SJ). The complete conditional probability density curve of future crop yields are illustrated 2) Two interval prediction evaluation metrics (prediction intervals coverage probability and prediction interval normalized average width) are used to assess the performance of the proposed QRF-SJ. 3) We demonstrated the superiority and feasibility of the proposed QRF-SJ model using groundnut and millet as case studies. 4) The feature/variable importance (a score that gives the effectiveness of each feature in predicting the yield of the crop) is presented. This gives information to agricultural stakeholders about the imporatant weather variable that affect the yield of the selected crops. The rest of the paper is organized as follows: section \ref{S:2} explains the theoretical background of quantile regression, random forest, quantile random forest, and kernel density estimation. The materials and methods are presented in Section \ref{S:3}. In section \ref{S:4}, the results and discussion of the proposed model using a case study of crop yield dataset is presented. The conclusion is outlined in section \ref{S:5}. 

\section{Theoretical Background}
\label{S:2}
This section provides a comprehensive explanation used in developing the probabilistic crop yield forecasting. Generally, three steps are used for the probabilistic crop yield forecasting. Firstly, the dataset is divided into a training and testing dataset. In the second phase, the training dataset is used to train the quantile regression forest (QRF) model. The QRF model is then used to predict the testing data on a selected quantile. In the final step, the probability density function are obtained by using kernel density estimation with Epanechnikov kernel function and SJ bandwidth selection. Our model has not been applied in other field of research or in the agriculture sector so far. It is therefore the first of its kind in literature.  

\subsection{Quantile Regression (QR)}
Conventional linear regression models make a summary of the average relation betweeen explanatory variables $X=[X_1, X_2, \cdots,X_k]'$ and a response variable $Y$ depending on the conditional mean function $\mathbb{E}(Y|X)$. It gives a partial estimate of the relationship, as it might be needed in recounting the relationship of distinct points in the conditional distribution of $Y$. Contrary to the conventional linear regression, QR gives the quantiles of the conditional distribution of $Y$ as a function of $X$ \citep{koenker1978regression}. That is, QR provides much detail information about the distribution of $Y$ than conventional linear regression model. By using QR, we can make a good inference on the distribution of the predicted values. Machine learning techniques that are based on quantile regression such as the quantile random forest have an extra advantage of being able to predict non-parametric distributions. A QR problem can be formulated as;
\begin{equation}
q_Y( {\tau}\mid X) = \mathbf{X}'_i \boldsymbol{\beta}_{\tau}
\label{QR_0} 
\end{equation}
where $q_Y({\tau} \mid \cdot)$ is the conditional $\tau_{th} (0 < \tau < 1)$ quantile of crop yield variables $Y$, $X$ are the explanatory variables, and $\beta_{\tau} = [\beta_{\tau}(0),\beta_{\tau}(1), \cdots \beta_{\tau}(k)]'$ is a vector of values of quantile $\tau$. By minimizing the loss function of a specific $\tau_{th}$ quantile, vector of values can be evaluated, 
\begin{equation}
\begin{aligned}
\underset{\beta}{\text{min}} \sum_{i=1}^{N}\rho_{\tau} (Y_i - \mathbf{X}'_i \boldsymbol{\beta})  =& \underset{\beta}{\text{min}} \Big[ \sum_{i:Y_i \ge \mathbf{X}'_i \boldsymbol{\beta}}^{} \tau \mid Y_i - \mathbf{X}'_i \boldsymbol{\beta} \mid  + \sum_{i:Y_i < \mathbf{X}'_i \boldsymbol{\beta}}^{} (1-\tau)\mid Y_i - \mathbf{X}'_i \boldsymbol{\beta} \mid \Big],\\
=& \underset{\beta}{\text{min}}\Big[ \sum_{i} \mid \tau - \mathbf{1}_{y_i < \mathbf{X}'_i \boldsymbol{\beta}} \mid (Y_i - \mathbf{X}'_i \boldsymbol{\beta}) \Big] 
\label{QR_02}
\end{aligned}
\end{equation}
Where $\mathbf{1}$ is the indicator function, $N$ is the size of the sample data, and $\mathbf{X}_i = (x_{1i}, x_{2i}, x_{3i}, \cdots, x_{ki})$ are the independent variables. 
Consider the distribution of a discrete random variable $Y_i$ with a less-than-well-behaved density, then the conditionbal density function at the $\tau_{th}$ quantile given $x_i$ is defined as 
\begin{equation}
q_{\tau}(x) = \inf\{y : F_{\tau} \big(y \mid X = x \big) \ge \tau \}
\label{QR_1}
\end{equation}
where $F_{\tau} \big(y \mid X = x \big)$ is the distribution function for $Y_i$ conditional on $X_i$. 

\subsubsection{Prediction Interval Construction}
PI is built from the conditional quantiles of the crop yield predicted from the QRF. Particularly, the $(1-\tau) \times 100\%$ PI for crop yield ($Y$) given the weather features ($\mathbf{X}$) is constructed as $PI(x) = [q_{\tau/2}(Y \mid \mathbf{X} = x), q_{1 - \tau/2}(Y \mid \mathbf{X} = x)]$. For instance, the $90\%$ PI for crop yield is calculated as 
\begin{equation*}
PI(x) = [q_{0.05}(Y \mid \mathbf{X} = x), q_{0.95}(Y \mid \mathbf{X} = x)].
\end{equation*}
That is, for $x$, the crop yield is within the interval of $PI(x)$ with a very high probability.

\subsection{Random Forest (RF)}
RF is a binary tree machine learning algorithm and a non-parametric method for regression and classification problems. The objective of RF is to predict the square integrable random response $Y \in \mathbb{R}$ by computing the regression function $c(x) = \mathbb{E}\big[Y | {\bf X} = {\bf x}\big]$. Assume a training dataset $D_{n}= \{\big({\bf X}_i, y_i\big)_{i=1}^{n}|{\bf X}_i \in \mathbb{R}^{M}, y \in \mathbb{R} \}$ of an observed dataset is randomly selected from an (unknown) probaility distribution $({\bf x}_i, y_i) \sim ({\bf X}, Y)$. We seek to use $D_{n}$ to build an estimate. Where ${n}$ is the total number of training samples and $M$ is the total number of features. 
\\
[1mm]
Suppose $\theta$ is the parameter that determines a specific splitting node of RF regression trees. Let $T(\theta)$ be the decision tree under consideration. Consider the conditional distribution of $Y$ given $X=x$ depending on the decision tree and the event that $x$ can be determined at a point on the decision tree $R$. If there is one and only one leaf node which satisfies $x$ and is represented as $\ell (x, \theta)$ for the decision tree $T(\theta)$, then the prediction of a single tree $T(\theta)$ for a point $x$ in the observed data is the average over the observed values in $\ell (x, \theta)$. The weight vector $w_n(x, \theta)$ for the total observation in $\ell (x, \theta)$ is given as
\begin{equation}
w_n(x,\theta) = \dfrac{\mathbf{1}_{\{X_n \in R_{\ell(x,\theta)}}\}}{\{p:X_p \in \Omega_{\ell(x, \theta)}\}}.
\label{rf_weight}
\end{equation}
Where $\sum_{i=1}^{n}w_n(x,\theta) = 1$, and the prediction of the single tree $Y \mid X=x$ is the weighted average of true observation $Y_i(i=1,2,\cdots,n)$,
\begin{equation}
\hat{\vartheta}(x) = \sum_{i=1}^{n} w_n(x, \theta)Y_n
\label{rf_weight_1}
\end{equation}
RF uses the average prediction of $k$ individual trees, each built with an independently and identically distributed (i.i.d.) vector $\theta_i\,\,i=1,2,3, \cdots,k$ to approximate $\mathbb{E}(Y \mid X=x)$. Denote $w_n(x)$ as the average of of $w_n(\theta)$ over the ensemble of trees, 
\begin{equation}
w_n(x) = \dfrac{1}{k} \sum_{i=1}^{K} w_n(x, \theta_i)
\end{equation}
Then, the prediction of RF is
\begin{equation}
\hat{\vartheta}(x) = \sum_{n=1}^{N}w_n(x)Y_n
\end{equation}
RF estimates the conditional mean of $Y$, given $X=x$, by weighting the sum of all the observations. The weight is larger when the conditional distribution of $Y$ given $X=X_n$, is identical to the conditional distribution of $Y$ given $X=x$ \citep{lin2006random}. 
\\
[1mm]
RF depends on some parameters for optimal performance. The number of tress (ntree) to grow and the number of variables that is sampled as candidates for each split (mtry). For regression problems, mtry = $\frac{M}{3}$, where $M$=number of features for prediction. Apart from using RF for quantile regression forest, we shall use RF for feature importance  and partial dependence plots (PDP). For the feature importance, we use the percentage mean decreasing accuracy (``\%IncMSE") to know the importance of each of the features in building the prediction model. The PDP illustrates how the RF model predictions are affected by each feature assuming the rest of the features in the RF model are controlled.  

\subsection{Quantile Random Forest (QRF)}
Conventional RF predict values in individual leaf node, which is considered as the sample mean in the leaf node. This can lead to biasness, that is extreme values in the data samples can be over- or under-estimated. To improve the accuracy of the prediction in the presence of extreme values in the sample dataset, the median can be used. Hence, the median is used for point prediction in QRF model.  
\\
QRF is a robust, non-linear, and non-parametric regression method based on random forests method for determining conditional quantiles \citep{meinshausen2006quantile}. QRF gives an approximation of the complete conditional distribution. Just like RF, QRF is a set of binary regression trees. However, for each leaf node of the tree, QRF evaluates the estimated distribution $F(y \mid X = x) = P(Y \le y) \mid X=x) = \mathbb{E}(1_{\{Y \le y\}} \mid X = x)$ as alternative to only the mean of $Y$ values in RF. Given a probability $p$, the quantile $q_\tau(X)$ is evaluated as $\hat{q}_\tau (X = x_{new}) = \inf \{y : \hat{F}(y \mid X=x_{new}) > \tau \}$. The quantiles provide a comprehensive information on the distribution of $Y$ as a function of the predictands ($X$) than only the conditional mean. For interval prediction, 
\begin{equation}
	\Big[q_{\tau_l}(X), q_{\tau_u}(X)\Big] = \Big[\inf\{y: \hat{F}(y \mid X =x) \ge \tau_l \}, \inf\{y:\hat{F}(y \mid X = x) \ge \tau_u\}\Big]
\end{equation}
where $\tau_l < \tau_u$ and $\tau_u - \tau_l = \alpha$, $\alpha$ is the probability that the predicted value fall within $y$ to lie in the interval $[q_{\tau_l}(X), q_{\tau_u}(X)]$. 
\\
[1mm]
We define an approximation to the acummulated conditional probability $\mathbb{E}(\mathbf{1}_{\{Y \le y\}} \mid X=x)$ by the weighted mean of all the observations of $\mathbf{1}_{\{Y \le y\}}$ as, 
\begin{equation}
F(y \mid X=x) = \sum_{n=1}^{N}w_n (x) \mathbf{1}_{\{Y_n \le y\}}, 
\label{QRF_1}
\end{equation}
where $w_n(x)$ is the same weights as in random forests. By plugging $\hat{F}\big(y \mid X =x\big)$ into \ref{QR_1}, the estimate $\hat{q}_{\tau}(x)$ of the conditional quantiles $q_{\tau}(x)$ are derived, 
\begin{equation}
\hat{q}_{\tau}(x) = \hat{F}^{-1}(\tau) = \inf \{y : \sum_{n=1}^{N}w_n (x) \mathbf{1}_{\{Y_n \le y\}} \ge p \}
\end{equation}

\subsection{Kernel density estimation  using Epanechnikov Kernel function}
Kernel density estimation (KDE) is a non-parametric method of estimating the probability density function (pdf) or regression functions. KDE is basically used for data smoothing. A Kernel density estimator at $x$ for an observed i.i.d. and  data ${\bf X} = (X_1, X_2, \cdots, X_n)$ drawn from an unknown distribution with an unknown density $f_X(x$), is
\begin{equation}
\hat{f} (x;b) = \dfrac{1}{Nb} \sum_{i=1}^{N} K \Big(\dfrac{X_i- x}{b} \Big)
\label{kde_1}
\end{equation}
where $N$ is the sample size, $K$ is the Kernel function, $h>0$ is the smoothing parameter also called bandwidth. The Kernel function is non-negative and is defined as $\int_{-\infty}^{\infty} K(x)dx=1$. Gaussian, Rectangular, Uniform, Cosine, Epanechnikov, and Quartic are but some common examples of kernel functions used in literatures. Different results are obtained depending on the type of Kernel function used. In this study, we use the Epanechnikov Kernel to build our QRF-SJ model.  
The Epanechnikov Kernel is defined as:  
\begin{equation}
K(u) = \begin{cases}
\frac{3}{4} (1 - u^2) \qquad \mid u \mid \le 1, \\
0 \qquad\qquad\qquad\,\,\,  \text{otherwise}. 
\end{cases}
\label{epanechnikov}
\end{equation}
Where $u = \frac{X_i - x}{b}$. The choice of the Epanechnikov kernel is motivated because it has the lowest (asymptotic) mean square error (MSE) \citep{epanechnikov1969non,wand1994kernel}.

\subsubsection*{Solve-The-Equation Plug-In Approach of Sheather and Jones (SJ)}
\begin{figure}[H]
	\centering
	\includegraphics[height=12cm,width=11.5cm]{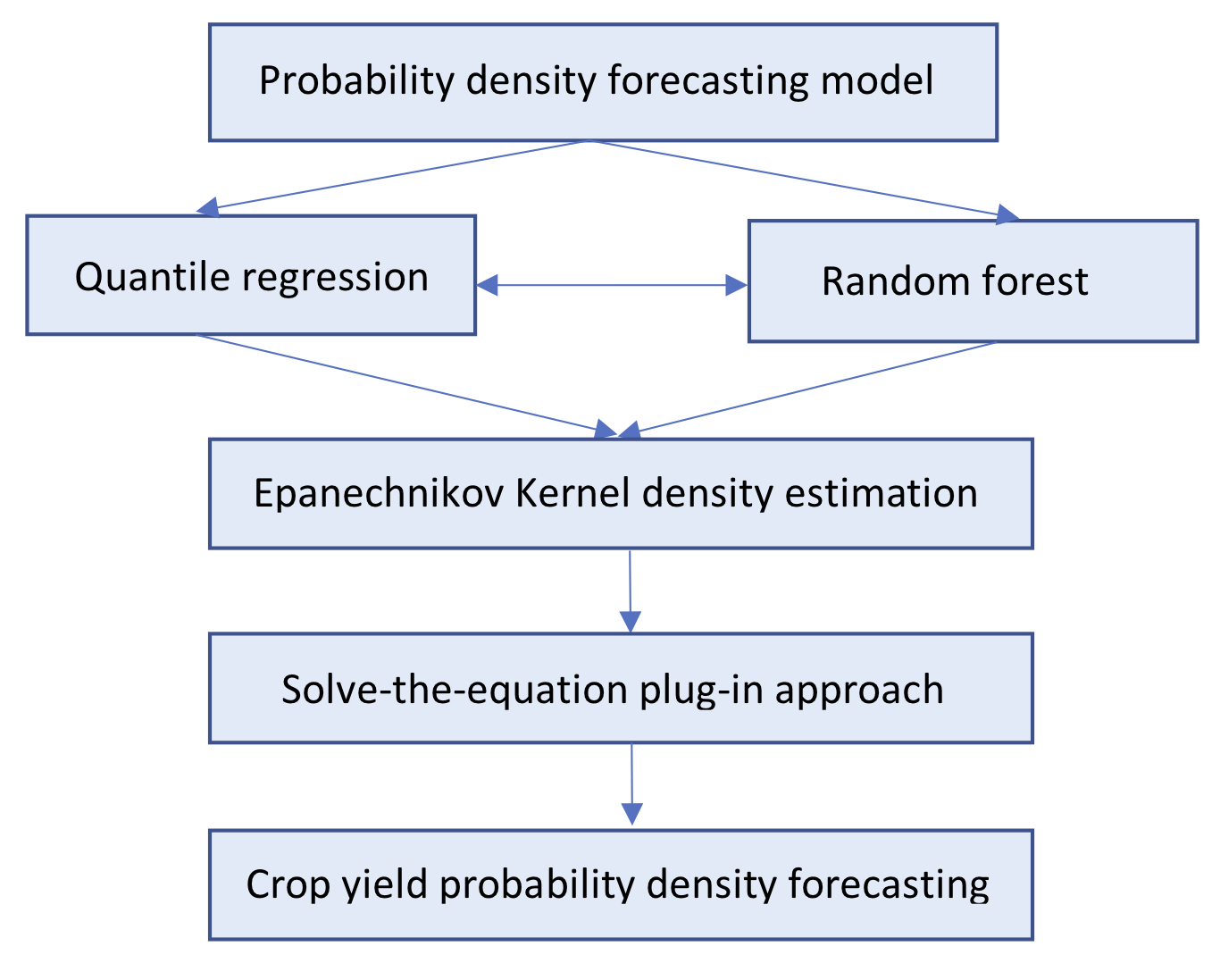}
	\caption{The block diagram of the proposed crop yield probability density forecasting method}	
	\label{blockdiagram}
\end{figure}

\begin{figure}[H]
	\centering
	\includegraphics[height=14cm,width=14cm]{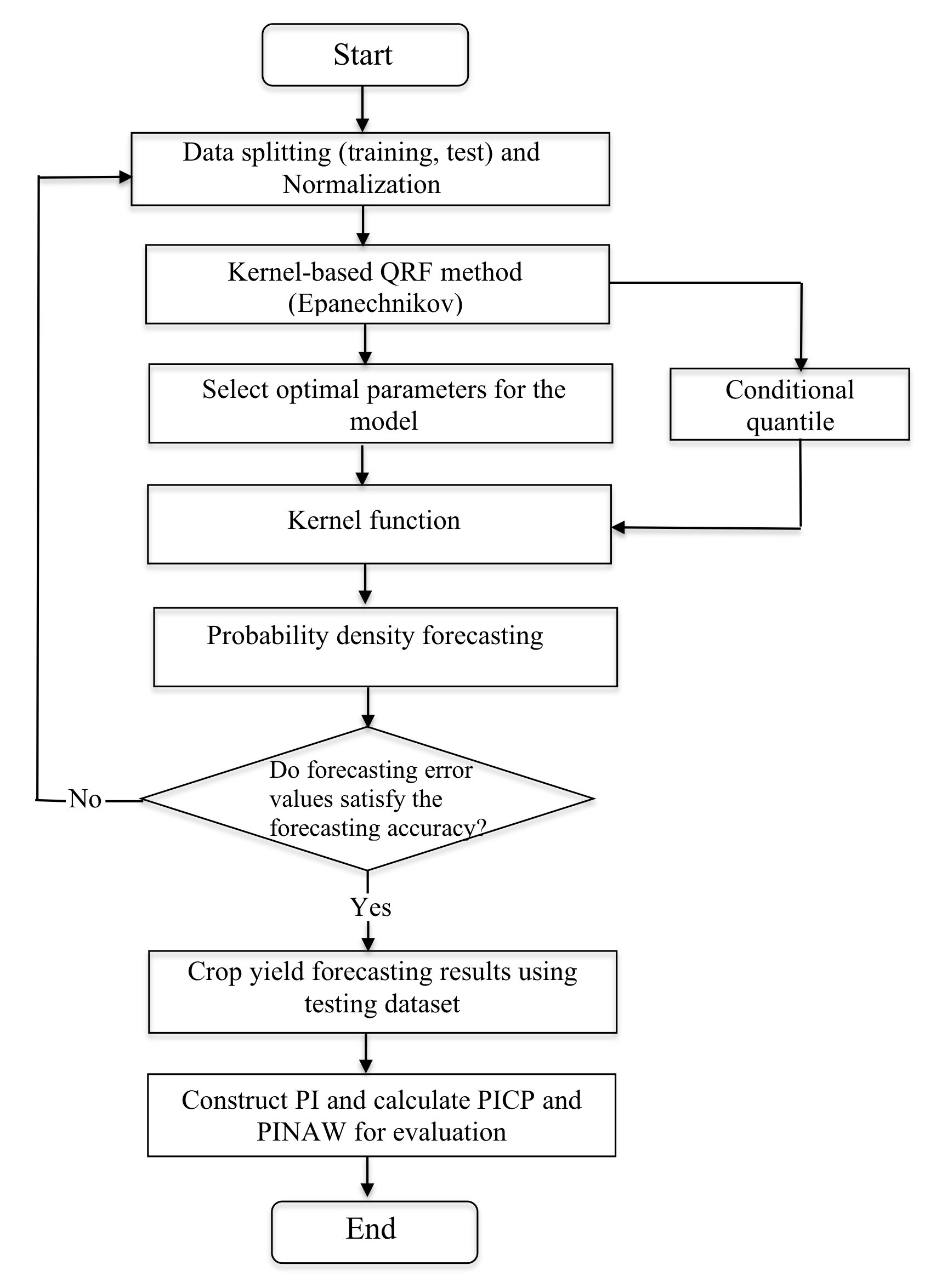}
	\caption{The flowchart of the structutre of the paper}	
	\label{flowchart}
\end{figure}

Bandwidth determines the smoothness of the kernel density plot and is comparable to the binwidth in a histogram. The selection of a proper bandwidth is the most difficult problem in obtaining a good KDE \citep{wand1994kernel}.
A larger value bandwidth value causes over smoothing and a very small bandwidth value causes under smoothing. To get better results of kernel density estimator, this paper uses Sheather and Jones (SJ) solve-the-equation (SJ) bandwidth selector \citep[see][]{jones1996brief, sheather1991reliable} for estimating the bandwidth parameter. 
\\
[1mm]
To quantify the accuracy of the kernel density estimator, the asymptotic mean integrated squared error (AMISE) is used. AMISE is an approximation of mean integrated squared error ( when $n \rightarrow \infty $ and $b=b(n) \rightarrow 0$) of $\hat{f}(x)$,  
\begin{equation}
AMISE(\hat{f}_b(x)) = (nb)^{-1}R(K) + b^4R(f'')  \Big( \int x^2K/2\Big)^2
\label{amise}
\end{equation}
where the notation $R(g) = \int g^2(x)dx$ for a function $g$, $\int x^2K = \int x^2K(x)dx$, and $f''$ is the second derivative of $f$. The first and second term in equation \ref{amise} are the integrated variance and integrated squared bias respectively. A very small $h$ results in a large integrated variance and a very large $h$ results in a large integrated squareed bias. 
\\
[2mm]
The flowchart of the proposed crop yield probability density forecasting method is clearly presented in figure \ref{blockdiagram}. Figure \ref{flowchart} shows flowchart of the structutre of the paper. 

\section{Materials and Methods }
\label{S:3}
\subsection{Evaluating point prediction errors}
We use root mean squared error (RMSE), mean absolute percentage error (MAPE), R-squared ($R^2$), and Bias to compare the performance of different forecasting models for point/deterministic forecasting. 
\\
RMSE (see equation \ref{RMSE}) estimates the residual betweeen the observed and the predicted values. The smaller the RMSE, the better the model. 
\begin{equation}
RMSE = \sqrt{ \dfrac{\sum_{i=1}^{N}(\hat{y}_i - y_i)^2}{N}} 
\label{RMSE}
\end{equation}
MAPE (see equation \ref{MAPE}) gives the average of the absolute percentage errors. An optimal model has the lowest MAPE. 
\begin{equation}
MAPE = \dfrac{1}{N} \sum_{i=1}^{N} \left| \dfrac{y_i - \hat{y}_i}{y_i} \right| \times 100\%
\label{MAPE}
\end{equation}
$R^2$ (see equation \ref{rsquared}) is a statistical measure that explains the variation of the observed to the predicted. Generally, the closer the value of $R^2$ to 1 (100\%), the better the model. 
\begin{equation}
R^2  = 1 - \dfrac{\sum_{i=1}^{N}(y_i - \hat{y}_i)^2}{\sum_{i=1}^{N}(y_i - \bar{y}_i)^2}
\label{rsquared}
\end{equation} 
Bias calculates the average amount by which the observe is greater than the predicted. For an unbiased model, the Bias should be closer to 0. Bias is computed as in equation (\ref{bias})
\begin{equation}
Bias  = \dfrac{1}{N}{\sum_{i=1}^{N}(y_i - \hat{y}_i)}
\label{bias}
\end{equation} 
$y_i$, $\hat{y}_i$, $\bar{y}_i$, $N$ are the observed/actual, predicted, mean, and the total number of dataset. The training data was used in training the models and the testing data was used in evaluating the performance of all the models. 

\subsection{Uncertainty of Prediction Intervals}
Different metrics are used to evaluate the prediction intervals for the results obtained from the probability density forecasting; prediction intervals coverage probability (PICP) and prediction interval normalized average width (PINAW). 
\\
PICP is the percentage of the testing data that fall in the interval specified by the upper bound $U_i$ and the lower bound $L_i$ of the prediction interval (PI). A larger PICP indicates that most of the forecasted data fall within the PI. Generally, the value of the PICP should be greater than the nominal confidence level. 
\begin{equation*}
PICP = \dfrac{1}{N}\sum_{i=1}^{N} c_i, 
\label{picp}
\end{equation*}
where $N$ is the total number of years over the period of forecasting and $c_i$ is a Boolean variable define as
\begin{equation*}
c_i = \begin{cases}
1, \qquad\qquad if \,\,\, y_i \in [L_i, U_i] \\0 \qquad\qquad\, if \,\,\, y_i \notin [L_i, U_i]
\end{cases}
\label{picp_c}
\end{equation*}
If the quality of the forecast depends only on the PICP, the coverage probability can be artificially improved by increasing the range between the upper and the lower bound. However, a larger interval width is empirically not informative. To better evaluate the quality of the PIs, the width of the PIs must be measured. A narrow PI gives more information to the forecaster than a wider PI. Therefore, a normalized metric PINAW which measures the average width of the PIs can be use. PINAW is defined as:
\begin{equation*}
PINAW = \dfrac{1}{NR} \sum_{i=1}^{t} (U_i - L_i)
\label{pinaw}
\end{equation*}
where $R$ is the range of the underlying targets (difference between minimum and maximum targets)

\section{Results and Discussion}
\label{S:4}
To demonstrate the feasibility and suitability of the proposed QRF-SJ method, historical crop yield{\footnote{crop yield is defined as the harvested production of a crop per unit of the harvested area and is measured in metric ton per hectare (t/ha).}} data for different crops in Tamale metropolitan{\footnote{The capital town of the Tamale metropolitan is Tamale, which also happens to be the regional capital of the Northern region.}} (a metropolis in the Northern region) were obtained from the Statistics, Research and Information Directorate (SRID) of the Ministry of Food and Agriculture, Ghana. The Northern region of Ghana of which the Tamale metropolitan is located is much drier{\footnote{This is because of its close  proximity to the Sahara, and the Sahel.}} as compared to the southern part of Ghana and agriculture contribute more than 75\% of the economic activities in the metropolis. The region is considered to be the food basket of Ghana. Cowpea, cassava, groundnut, maize, millet, sorghum, rice, and yam are the major crops grown in this region. For the purpose of this research, two of the crops (groundnut and millet) are selected as a case study to evaluate the performance of our proposed model. These selected crops can be used as a proxy to create an area-yield index insurance instrument for the insurance sector. Due to the unavailability of historical data in the metropolis, the selected crop yield data was taken from 2000 to 2016. The growing seasons of groundnut are March-May and September-October, while the growing season of millet is from June to October \cite{fao1} . Station based daily sunlight, humidity, precipitation, minimum temperature, maximum temperature and average temperature from 2000 to 2016 are obtained from the Ghana Meteorological Service. The k-nearest neighbors (KNN) algorithm was used for imputing missing data points in the datasets. KNN locates the $k$ closest neighbors to the observed dataset with the missing data point and imputes the data point based on the non-missing data points in the neighbors. The average of the climatological data over the cropping season of groundnut and millet are computed as the growing season climatological factors. 
\\
Because of the scaling sensitivity of the inputs fed into most forecasting techniques, the variables for the inputs are set into an identical scale. The scale used is the min-max normalization which was set to be in the interval $[0,1]$. The normalization is given as:
\begin{equation*}
I_{NORM} = \dfrac{I-I_{MIN}}{I_{MAX} - I_{MIN}}
\end{equation*}
where   $I_{NORM}$  is the normalized numerical value;   $I_{MIN}$, $I_{MAX}$ is the minimum and maximum values of the inputs respectively.
\\
[1mm]
To validate the model, we divided the dataset into a training (80\%) and testing (20\%) dataset. That is, the dataset from 2000-2013 are selected as the training data and 2014-2016 are selected as the testing data.

 $R$ statistical software was used in implementing all the forecasting approaches.  
\begin{table}[H]
	\centering
	\begin{tabular}{cccccccccc}		\hline 
		Crop & Mean   &  Std  & Min   & Max &  Skewness  \\ 
		\hline 
		Groundnut yield (t/ha) &   1.22  & 0.50 & 0.50 & 1.90  & 0.05  \\
		Millet yield (t/ha)&   1.19  & 0.30 & 0.72 & 1.70  & 0.10  \\
		\hline 
	\end{tabular} 
	\caption{Summary Statistics of Groundnut and Millet Yield}
	\label{summary_statistics}
\end{table}

\subsection{Empirical results and analysis of the Models}
\begin{figure}[H]
	\centering~
\includegraphics[height=8.1cm,width=14cm]{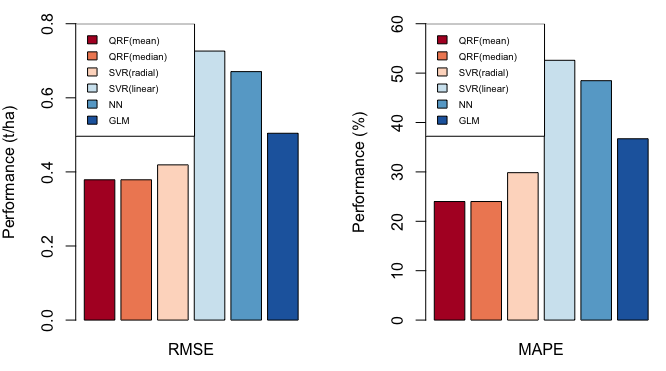}
\includegraphics[height=8.1cm,width=14cm]{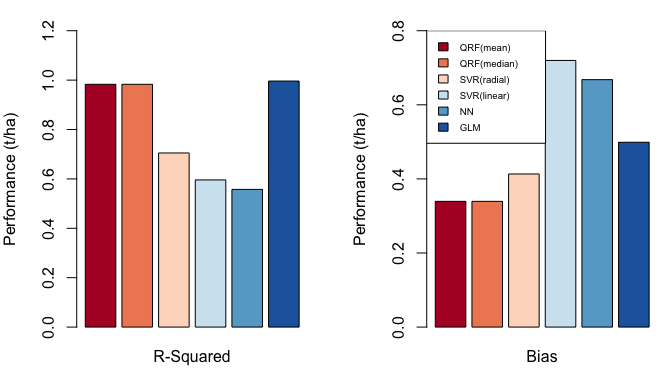}
	\caption{Bar chart of RMSE, MAPE, R-Squared, and Bias of observed and predicted groundnut yield by QRFs, SVR(radial), SVR(linear), NN, and GLM using testing data. QRF (mean): mean prediction of crop yield given weather features using QRF; QRF (median): median prediction of crop yield response using QRF; SVR(radial): prediction of crop yield using SVR radial basis kernel function; SVR(linear): prediction of crop yield using SVR linear kernel.}	
	\label{groundnut_comparison_MLmethod}
\end{figure}

\begin{figure}[H]
	\centering~
	\includegraphics[height=8.1cm,width=14cm]{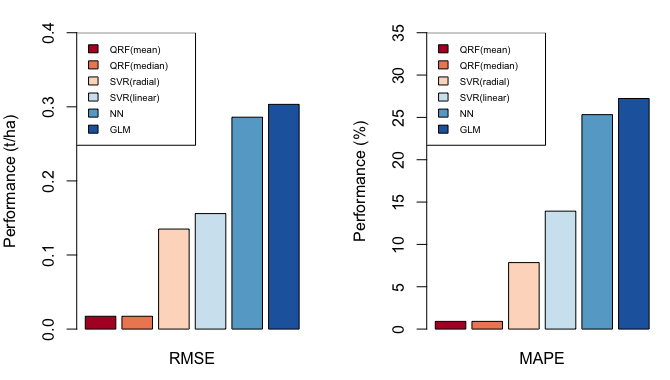}
	\includegraphics[height=8.1cm,width=14cm]{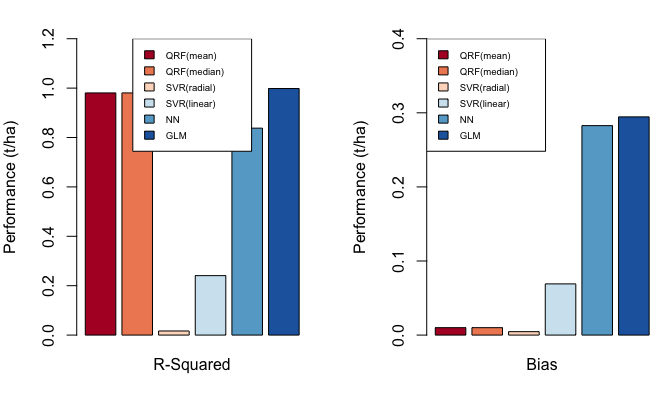}
	\caption{Bar chart of RMSE, MAPE, R-Squared, and Bias of observed and predicted millet yield by QRFs, SVR(radial), SVR(linear), NN, and GLM using testing data. QRF (mean): mean prediction of crop yield given weather features using QRF; QRF (median): median prediction of crop yield response using QRF; SVR(radial): prediction of crop yield using SVR radial basis kernel function; SVR(linear): prediction of crop yield using SVR linear kernel.}	
	\label{millet_comparison_MLmethod}
\end{figure}

In order to prove the superiority of the QRF, it is compared with some popular forecasting techniques like Radial Basis Neural Network (NN), Generalized Linear Model (GLM), Support Vector Regression with linear (SVR (linear)) and radial basis (SVR (radial)) kernel function. The hyperparameters used in tuning the ML techniques is detailed in appendices \ref{appendix_groundnut} and \ref{appendix_millet}. 
\\
The evaluation metrics (RMSE, MAPE, $R^2$, and Bias) of these methods are given in Table \ref{benchmarks_evaluation_metrics}.  Figures \ref{groundnut_comparison_MLmethod} and \ref{millet_comparison_MLmethod} show the visual performance of the evaluation metrics of the forecasting techniques. Comparative to the four benchmark methods, it is evident that QRF (both mean and median prediction) performed better in terms of RMSE, MAPE and Bias when predicting the yield of groundnut and millet for the testing data. GLM outperformed all the other methods when using $R^2$ as the performance metrics for both groundnut and millet.  It can be observed that the QRF median prediction is the same as the mean prediction. Quantitatively, the RMSE, MAPE, $R^2$ and Bias of QRF (mean and median) for groundnut are $0.3787$\,t/ha, $24.0026\%$, $0.9830$, and $0.3394$\,t/ha respectively. The RMSE, MAPE, $R^2$ and Bias of QRF (mean and median) for millet are $0.0173$\,t/ha, $0.9090\%$, $0.9805$ and $0.01$\,t/ha respectively. Generally, QRF is optimal in predicting the yield of all the two crops. For this reason, we conclude that QRF is the best forecasting technique for predicting the yield of the two crops as compared to the other benchmark forecasting techniques. This motivated us to use QRF for the probability density forecasting. 

\begin{table}[H]
	\renewcommand{\arraystretch}{1.5}
	\centering	
	\resizebox{\columnwidth}{!}{%
		\begin{tabular}{>{\bfseries}c*{17}{c}} 
			\toprule
			\multirow{2}{*}{\bfseries Method} &
			\multicolumn{5}{c}{\bfseries  Groundnut} &
			\multicolumn{5}{c}{\bfseries \qquad Millet} \\\cmidrule(lr){3-6} \cmidrule(lr){8-11}
			&& \textbf{RMSE (t/ha)} & \textbf{MAPE (\%)} &  \textbf{$R^2$} & \textbf{Bias (t/ha)} & & 
			\textbf{RMSE (t/ha)} & \textbf{MAPE (\%)} & \textbf{$R^2$} & \textbf{Bias(t/ha)}
			\\ \hline
			QRF(mean) && {\bf 0.3787} & {\bf 24.0026} & 0.9830 & {\bf 0.3394} && {\bf 0.0173} & {\bf 0.9090} & 0.9805 & {\bf 0.0100}   
			\\ 
			QRF(median) && {\bf 0.3787} & {\bf 24.0026} & 0.9830 & {\bf 0.3394} && {\bf 0.0173} & {\bf 0.9090} & 0.9805 & {\bf 0.0100}            
			\\
			SVR (radial) && 0.4190 & 29.8431 & 0.7050 & 0.4133 && 0.1035 & 7.8520 & 0.0166 & 0.0046  
			\\
			SVR (linear) && 0.7262 & 52.5940 & 0.5959 & 0.7198 && 0.1559 & 13.9280 & 0.2407 & 0.0691        
			\\
			NN && 0.6707 & 48.4656 & 0.5575 & 0.6679 && 0.2860 & 25.3258 & 0.8379 & 0.2827           
			\\
			GLM && 0.5043 & 36.7089 & {\bf 0.9962} & 0.4989 && 0.3033 & 27.2320 & {\bf 0.9983} & 0.2945             
			\\
			\bottomrule
		\end{tabular}
	}
	\caption{Evaluation metrics (RMSE, MAPE, R-squared, Bias) of point prediction via QRFs, SVR(radial), SVR(linear), NN, and GLM using testing data.}
	\label{benchmarks_evaluation_metrics}
\end{table}

\begin{figure}[H]
	\centering
	\includegraphics[height=8cm,width=15cm]{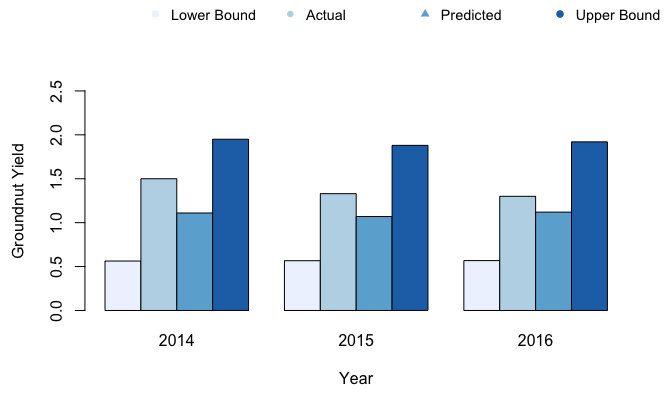}
	\includegraphics[height=7cm,width=15cm]{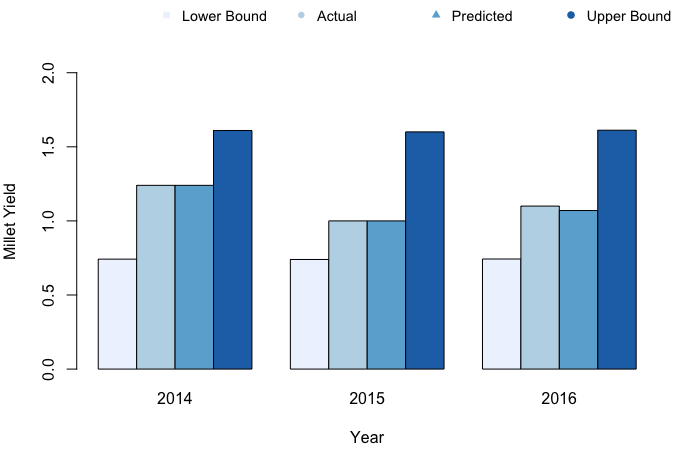}
	\caption{Prediction results and prediction intervals for the yield of groundnut and millet.}	
	\label{millet_groundnut_actual_pred_den_barplot}
\end{figure}

Table \ref{groundnut_millet_actual_pred_den_chart} presents the values of the prediction results  and prediction intervals based on QRF-SJ forecast for testing data. From this table, it is clear that each prediction has its own uncertainty bound. Figure \ref{millet_groundnut_actual_pred_den_barplot} gives the point forecast and the prediction interval of groundnut and millet respectively. To assess the performance of point forecast for the model, the performance metrics over the predicted period for the two crops are presented in table \ref{evaluation_metrics_point}. It is clear from figure \ref{millet_groundnut_actual_pred_den_barplot} that the target values lie within the prediction interval. We can therefore conclude that the proposed model captures the uncertainty of the two crop yields accurately. 

\begin{table}[H]
	\renewcommand{\arraystretch}{1.5}
	\centering	
	\resizebox{\columnwidth}{!}{%
		\begin{tabular}{>{\bfseries}c*{17}{c}} 
			\toprule
			\multirow{2}{*}{\bfseries Year} &
			\multicolumn{5}{c}{\bfseries  Groundnut} &
			\multicolumn{5}{c}{\bfseries \qquad Millet} \\\cmidrule(lr){3-6} \cmidrule(lr){8-11}
			&& \textbf{Lower bound} & \textbf{Observed} &  \textbf{Predicted} & \textbf{Upper bound} & & 
			\textbf{Lower bound} & \textbf{Observed} & \textbf{Predicted} & \textbf{Upper bound}
			\\ \hline
			2014 && 0.5630 & 1.5000 & 1.1100 & 1.9500 && 0.7420 & 1.2400 & 1.2400 & 1.6101        
			\\
			2015 && 0.5667 & 1.3300 & 1.0700 & 1.8800 && 0.7397 & 1.0000 & 1.0000 & 1.6007           
			\\
			2016 && 0.5680 & 1.3000 & 1.1200 & 1.9200 && 0.7429 & 1.1000 & 1.0700 & 1.6120             
			\\
			\bottomrule
		\end{tabular}
	}
	\caption{Prediction results and prediction intervals for the yield of groundnut and millet for testing data}
	\label{groundnut_millet_actual_pred_den_chart}
\end{table}

\begin{table}[H]
	\resizebox{\columnwidth}{!}{%
		\renewcommand{\arraystretch}{1.3}
		\centering
		\begin{tabular}{cccccccc}		\hline 
			Crop & {\bfseries Confidence level (\%)} & {\bfseries MAPE (\%)} & {\bfseries RMSE (t/ha)} & {\bfseries Bias (t/ha)}  \\ 
			\hline 
			Groundnut & 90 & 27.50 & 0.2895 & 0.2399 \\
			Millet &   90  & $1 \times 10^{-14}$ & 0.0072 & 0.0147 \\
			\bottomrule 
		\end{tabular}
	}
	\caption{Evaluation metrics of point prediction via QRF-SJ for testing data}
	\label{evaluation_metrics_point}
\end{table}

\begin{figure}[H]
	\centering
	\includegraphics[height=7cm,width=14cm]{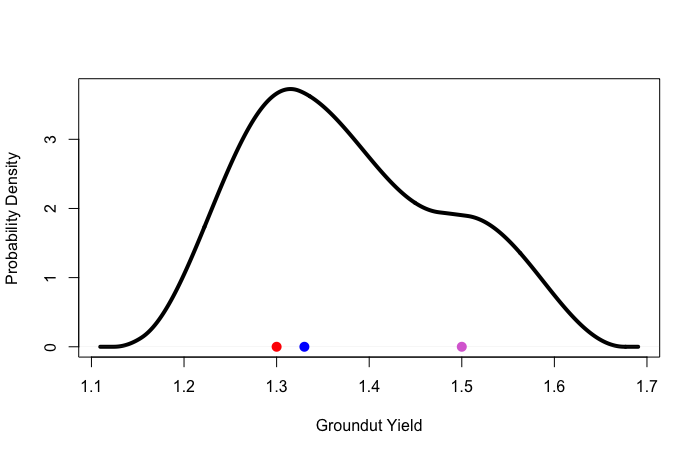}
	\caption{probability density curve based on QRF-SJ from 2014-2016, the dots on the x-axis represents the actual values of the crop yield.}	
	\label{groundnut_density_curve}
\end{figure}

\begin{figure}[H]
	\centering
	\includegraphics[height=7cm,width=14cm]{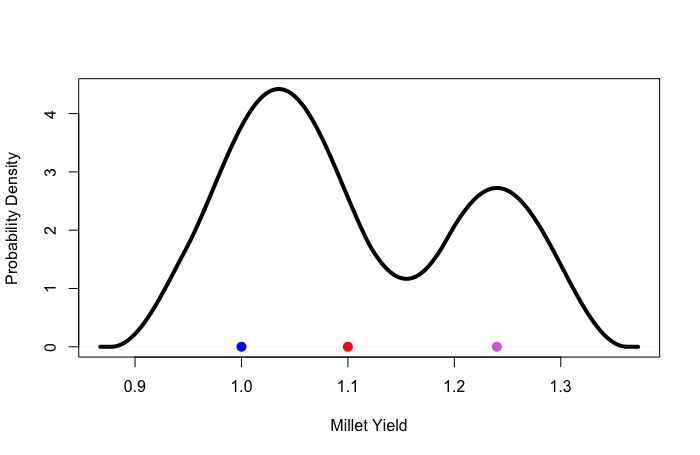}
	\caption{probability density curve based on QRF-SJ from 2014-2016, the dots on the x-axis represents the actual values of the crop yield.}	
	\label{millet_density_curve}
\end{figure}

To show the satisfactory performance of the proposed QRF-SJ model, PICP and PINAW are used as the evaluation metrics. The performance of the model is presented in Table \ref{evaluation_metrics_PI}. It is clear from the table that, all observed values were captured by the confidence interval. This is very important for effective decision-making process. For both groundnut and millet, the PICP was evaluated to be 100\%. The PINAW of groundnut is however smaller than that of millet. Considering the high variance of the yield of groundnut in table \ref{summary_statistics}, the probabilistic performance of our proposed method is sufficient. 
\\
A visual representation of the probability density curve for the predictions based on QRF-SJ for the yield of groundnut and millet is presented in figures \ref{groundnut_density_curve} and \ref{millet_density_curve} respectively. The actual crop yield for the specific year are presented in orchid, blue, and red dots for 2014, 2015, and 2016 respectively. In figures \ref{groundnut_density_curve} and \ref{millet_density_curve}, the respective actual crop yields of groundnut and millet for each of the predicted year fall within the predicted region of the forecast distribution. The probability density curve gives the complete probability distribution of the future crop yield and hence the uncertainty of the forecasting can be quantified. 

\begin{table}[H]
	\renewcommand{\arraystretch}{1.3}
	\centering
	\begin{tabular}{cccccccc}		\hline 
		Crop & {\bfseries Confidence level (\%) } & {\bfseries PICP (\%)} & {\bfseries PINAW (\%)}   \\ 
		\hline 
		Groundnut & 90 & 100 & 12.65  \\
		Millet &   90  & 100 & 16.76  \\
		\bottomrule 
	\end{tabular} 
	\caption{Prediction Interval evaluation indices of QRF-SJ method}
	\label{evaluation_metrics_PI}
\end{table}

\subsection{Feature Importance}
\begin{figure}[H]
	\centering
	\includegraphics[height=8cm,width=16cm]{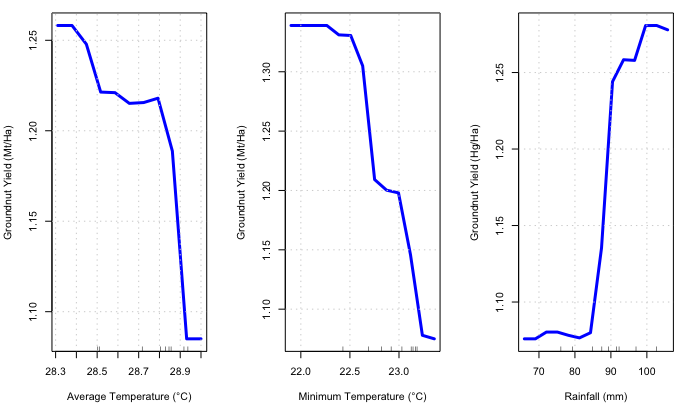}
	\caption{Partial dependence plot of groundnut for the top 3 ranked predictor variable from variable importance measures of Random Forests models using the testing data}	
	\label{groundnut_partial_plot}
\end{figure}

\begin{figure}[H]
	\centering
	\includegraphics[height=7cm,width=13cm]{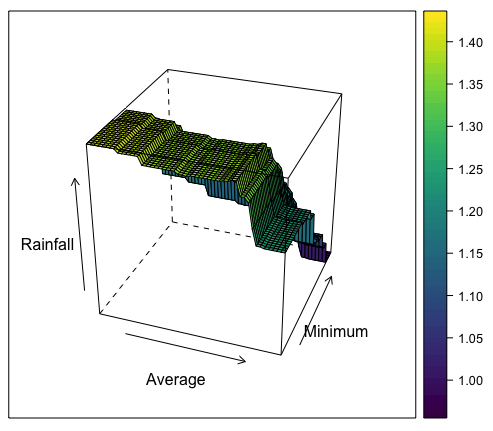}
	\caption{Groundnut: A 3-D surface Partial dependence of Rainfall on Average and Minimum Temperature based on a random forest using the testing data.}
	\label{groundnut_3_D_surface}
\end{figure}

\begin{figure}[H]
	\centering 
	\includegraphics[height=8cm,width=16cm]{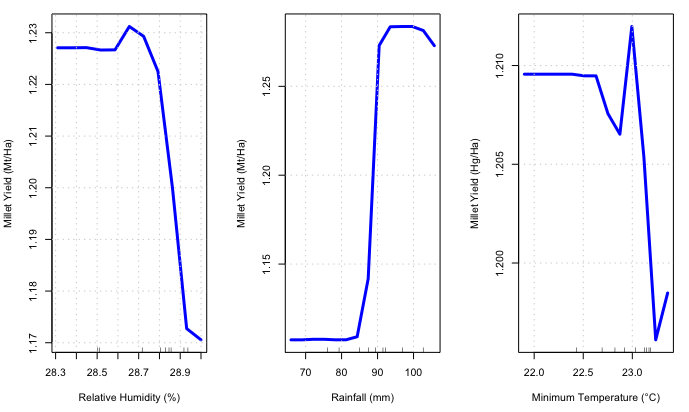}
	\caption{Partial dependence plot of millet for the top 3 ranked predictor variable from variable importance measures of Random Forests models using the testing data.}	
	\label{millet_partial_plot}
\end{figure}

\begin{figure}[H]
	\centering
	\includegraphics[height=7cm,width=13cm]{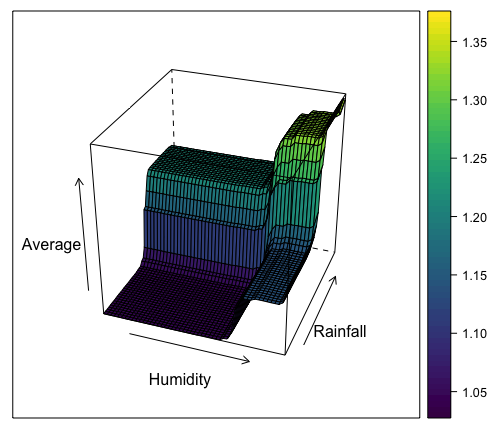}
	\caption{Millet: A 3-D surface Partial dependence of Average temperature on humidity and rainfall based on a random forest using the testing data.}
	\label{millet_3_D_surface}
\end{figure}

The level of feature importance measures according to random forest is shown in Table \ref{variable_importance}. The higher the percentage increase in mean square error (\%IncMSE) of a feature, the higher the importance of that feature in the prediction model. 
\\
From the feature importance measure in Table \ref{variable_importance}, average temperature, minimum temperature, and rainfall are the three most important features among the six features that affect the yield of groundnut. Humidity, rainfall, and average temperature are the top three features that influences the yield of millet. The amount of sunshine does not have a lot of effect on the yield of groundnut. Maximum temperature do not have a lot of effect on the yield of both crops. Generally, average temperature do have a lot of effect on the yield of these two crops. The partial dependence plots (PDP) in Figures \ref{groundnut_partial_plot} and \ref{millet_partial_plot} show the marginal effect of the top three important features of the yield of groundnut and millet respectively. 
\\
From the PDP of groundnut, an increase in the amount of average and minimum temperatures result in a decrease in the yield of groundnut. However, an increase in the amount of rainfall increases the yield of groundnut. An increasing relative humidity and minimum temperatures decreases the yield of millet. In both crops, the yield is minimum when the minimum temperature is around 23.2$^\circ$C. From the PDPs, it can be concluded that the yield of groundnut and millet increases as the amount of rainfall increases to about 100mm. The yield of groundnut and millet decreases as the minimum temperature increases from about 22$^\circ$C to 23.2$^\circ$C. Figures \ref{groundnut_3_D_surface} and \ref{millet_3_D_surface} show the three-dimensional (3-D) partial dependence of the top 3 ranked features from feature importance measures of Random Forests model for groundnut and millet respectively. From table \ref{variable_importance}, it is clear that all the six weather features do influence the yield of all the two crops. 

\begin{table}[H]
	\renewcommand{\arraystretch}{1.3}
	\centering
	\begin{tabular}{>{\bfseries}c*{11}{c}} 
		\toprule
		\multirow{2}{*}{\bfseries Feature} & 
		\multicolumn{2}{c}{\bfseries  Groundnut} &
		\multicolumn{3}{c}{\bfseries \qquad Millet} \\\cmidrule(lr){2-3} \cmidrule(lr){5-6}
		& \textbf{Rank} & \textbf{\%IncMSE}  & &\textbf{Rank} & \textbf{\%IncMSe}  \\ \hline
		Sunshine & 6 & 0.101 && 4 & 0.120         
		\\ 
		Humidity & 4 & 0.144 && 1 & 0.210              
		\\
		Rainfall & 3 & 0.162 && 2 & 0.198              
		\\
		AvgT     & 1 & 0.175 && 3 & 0.121              
		\\
		MaxT     & 5 & 0.125 && 5 & 0.114             
		\\
		MinT     & 2 & 0.171 && 6 & 0.115           
		\\
		\bottomrule
	\end{tabular}
	\caption{Rank corresponds to variable importance measure determined by Random Forest (RF) model for each crop dataset. AvgT, MaxT, MinT represent average, maximum and minimum temperature respectively.}
\label{variable_importance}	
\end{table}

\subsection{Discussion}
\subsubsection{QRF-SJ method}
Due to the variability of the yield of crops as a result of climate variability \cite{alexandrov2000impact}, a method QRF-SJ that can effectively handle this effect is proposed using quantile random forest and kernel density estimation. QRF does not only outperform other benchmark methods like SVR, NN, and GLM in point forecasting but also gives useful probability distribution when combined with kernel density estimation. Probability density forecasting has the ability to represent uncertainty of crop yields as the probability distributions around a prediction interval. Forecasting the distributions serves as an indicator for the accuracy of forecast and provides important information for decision making.
\\
Crop yield forecasting is an integral factor in precision farming and can promote the expansion of the agriculture sector. On the part of the government, this forecasting model will aid in effective decision planning. Hence, food shortages can be avoided and if possible governments can arrange for food imports rather than seeking for emergency food assistance. The proposed forecasting model can be used for trade development policies and other humanitarian assistance connected to food security. For the industry players like the insurance and financial sector, crop yield forecasting helps in measuring crop loss' in advance. Consequently, fair premium rates and pricing of agricultural index insurance and weather derivatives can be determined. The farmer however is able to measure the future uncertainties of the farm produce and make effective plans for a set of possible outcomes and in particularly precision farming.

\subsubsection{Feature Importance}
The feature importance gave a score of the importance of each weather variable in building the QRF model. By identifying the contributions of these weather variables to crop yields and applying effective forecasting models, the farmer for instance will be able to detect the specific weather variable that affects the yield of the crops and can adapt the appropriate risk management strategies. The insurance sector is able to realize the specific weather variable which correlates with the actual farm yield. Using the feature importance, insurance companies can sell different weather-index insurance products to farmers depending on the weather variable that affect the yield of their crops. Payouts of this weather insurance will be triggered based on the level for which the weather variable affects the yield of the crop. Base on feature importance, insurers and re-insurers companies are able to price their insurance product base on the specific climatic factors that affect the yield of a specific crop. They can also merge the most important climatic factors (compouns events) in their pricing models. 

\section{Conclusion}
\label{S:5}
Crop yield forecasting that considers the uncertainty of weather as a result of changes in climate is of crucial importance for efficient decision-making process in the agricultural sector. In this paper, a hybrid crop yield probability density forecasting method that has the potential to draw complete conditional probability density curve of future crop yields is presented. In the method, quantile regression forest is used to build the nonlinear quantile regression forecasting model and to capture the nonlinear relationship between the weather variables and crop yields. Epanechnikov kernel function and solve-the equation plug-in approach of Sheather and Jones are employed in the method to construct the probability density forecasting curve. Prediction interval coverage probability and prediction interval normalized average width are used to evaluate the quality of the prediction intervals constructed by QRF-SJ. 
The performance and accuracy of the QRF-SJ crop yield forecasting model are evaluated using two real dataset (groundnut and millet yields) as case studies. The numerical results gave a 100\% PICPs and narrow PINAWs. All the observed groundnut and millet crop yield values are located in the probability density curves.  The results show the superiority and feasibility of the proposed QRF-SJ model in forecasting the yield of crops in the midst of of weather uncertainties. Using the feature importance, agricultural stakeholders will be able to detect the major weather variables that affect the yield of crops and can make pragmatic interventions to curtail any uncertainties.


\section*{Acknowledgements}
The first author wishes to thank African Union and Pan African University, Institute for Basic Sciences Technology and Innovation, Kenya, for their financial support for this research.

\bibliographystyle{model1-num-names}
\bibliography{sample.bib}

\begin{thebibliography}{36}
\expandafter\ifx\csname natexlab\endcsname\relax\def\natexlab#1{#1}\fi
\providecommand{\bibinfo}[2]{#2}
\ifx\xfnm\relax \def\xfnm[#1]{\unskip,\space#1}\fi
\bibitem[{Rana and Rana(2014)}]{rana2014}
\bibinfo{author}{S.~S. Rana}, \bibinfo{author}{R.~S. Rana},
  \bibinfo{title}{ADVANCES IN CROP GROWTH AND PRODUCTIVITY},
  \bibinfo{type}{Technical Report}, Publication of the Department of Agronomy,
  CSK Himachal Pradesh Krishi Vishvavidyalaya, Palampur, India,
  \bibinfo{year}{2014}.
\bibitem[{Brevik(2013)}]{brevik2013potential}
\bibinfo{author}{E.~Brevik},
\newblock \bibinfo{title}{The potential impact of climate change on soil
  properties and processes and corresponding influence on food security},
\newblock \bibinfo{journal}{Agriculture} \bibinfo{volume}{3}
  (\bibinfo{year}{2013}) \bibinfo{pages}{398--417}.
\bibitem[{Chen et~al.(2004)Chen, McCarl, and Schimmelpfennig}]{chen2004yield}
\bibinfo{author}{C.-C. Chen}, \bibinfo{author}{B.~A. McCarl},
  \bibinfo{author}{D.~E. Schimmelpfennig},
\newblock \bibinfo{title}{Yield variability as influenced by climate: A
  statistical investigation},
\newblock \bibinfo{journal}{Climatic Change} \bibinfo{volume}{66}
  (\bibinfo{year}{2004}) \bibinfo{pages}{239--261}.
\bibitem[{Isik and Devadoss(2006)}]{isik2006analysis}
\bibinfo{author}{M.~Isik}, \bibinfo{author}{S.~Devadoss},
\newblock \bibinfo{title}{An analysis of the impact of climate change on crop
  yields and yield variability},
\newblock \bibinfo{journal}{Applied Economics} \bibinfo{volume}{38}
  (\bibinfo{year}{2006}) \bibinfo{pages}{835--844}.
\bibitem[{Southworth et~al.(2000)Southworth, Randolph, Habeck, Doering,
  Pfeifer, Rao, and Johnston}]{southworth2000consequences}
\bibinfo{author}{J.~Southworth}, \bibinfo{author}{J.~Randolph},
  \bibinfo{author}{M.~Habeck}, \bibinfo{author}{O.~Doering},
  \bibinfo{author}{R.~Pfeifer}, \bibinfo{author}{D.~G. Rao},
  \bibinfo{author}{J.~Johnston},
\newblock \bibinfo{title}{Consequences of future climate change and changing
  climate variability on maize yields in the midwestern united states},
\newblock \bibinfo{journal}{Agriculture, Ecosystems \& Environment}
  \bibinfo{volume}{82} (\bibinfo{year}{2000}) \bibinfo{pages}{139--158}.
\bibitem[{Asfaw and Lipper(2016)}]{asfaw2016managing}
\bibinfo{author}{S.~Asfaw}, \bibinfo{author}{L.~Lipper},
  \bibinfo{title}{Managing climate risk using climate-smart agriculture},
  \bibinfo{publisher}{FAO}, \bibinfo{year}{2016}.
\bibitem[{Gornall et~al.(2010)Gornall, Betts, Burke, Clark, Camp, Willett, and
  Wiltshire}]{gornall2010implications}
\bibinfo{author}{J.~Gornall}, \bibinfo{author}{R.~Betts},
  \bibinfo{author}{E.~Burke}, \bibinfo{author}{R.~Clark},
  \bibinfo{author}{J.~Camp}, \bibinfo{author}{K.~Willett},
  \bibinfo{author}{A.~Wiltshire},
\newblock \bibinfo{title}{Implications of climate change for agricultural
  productivity in the early twenty-first century},
\newblock \bibinfo{journal}{Philosophical Transactions of the Royal Society B:
  Biological Sciences} \bibinfo{volume}{365} (\bibinfo{year}{2010})
  \bibinfo{pages}{2973--2989}.
\bibitem[{Gyamerah et~al.(2018)Gyamerah, Ngare, and Ikpe}]{gyamerah2018regime}
\bibinfo{author}{S.~A. Gyamerah}, \bibinfo{author}{P.~Ngare},
  \bibinfo{author}{D.~Ikpe},
\newblock \bibinfo{title}{Regime-switching temperature dynamics model for
  weather derivatives},
\newblock \bibinfo{journal}{International Journal of Stochastic Analysis}
  \bibinfo{volume}{2018} (\bibinfo{year}{2018}).
\bibitem[{Delinc{\'e} et~al.(2017)}]{delince2017recent}
\bibinfo{author}{J.~Delinc{\'e}}, et~al.,
\newblock \bibinfo{title}{Recent practices and advances for amis crop yield
  forecasting at farm and parcel level: a review.},
\newblock \bibinfo{journal}{Recent practices and advances for AMIS crop yield
  forecasting at farm and parcel level: a review.}  (\bibinfo{year}{2017}).
\bibitem[{Shi et~al.(2013)Shi, Tao, and Zhang}]{shi2013review}
\bibinfo{author}{W.~Shi}, \bibinfo{author}{F.~Tao}, \bibinfo{author}{Z.~Zhang},
\newblock \bibinfo{title}{A review on statistical models for identifying
  climate contributions to crop yields},
\newblock \bibinfo{journal}{Journal of geographical sciences}
  \bibinfo{volume}{23} (\bibinfo{year}{2013}) \bibinfo{pages}{567--576}.
\bibitem[{Michler et~al.(2015)Michler, Viens, Shively et~al.}]{michler2015risk}
\bibinfo{author}{J.~D. Michler}, \bibinfo{author}{F.~G. Viens},
  \bibinfo{author}{G.~E. Shively}, et~al.,
\newblock \bibinfo{title}{Risk, agricultural production, and weather index
  insurance in village south asia},
\newblock in: \bibinfo{booktitle}{International Association of Agricultural
  Economists 2015 International Conference of Agricultural Economists, Milan,
  Italy, August}, pp. \bibinfo{pages}{8--14}.
\bibitem[{Lobell and Burke(2010)}]{lobell2010use}
\bibinfo{author}{D.~B. Lobell}, \bibinfo{author}{M.~B. Burke},
\newblock \bibinfo{title}{On the use of statistical models to predict crop
  yield responses to climate change},
\newblock \bibinfo{journal}{Agricultural and Forest Meteorology}
  \bibinfo{volume}{150} (\bibinfo{year}{2010}) \bibinfo{pages}{1443--1452}.
\bibitem[{Choudhury and Jones(2014)}]{choudhury2014crop}
\bibinfo{author}{A.~Choudhury}, \bibinfo{author}{J.~Jones},
\newblock \bibinfo{title}{Crop yield prediction using time series models},
\newblock \bibinfo{journal}{Journal of Economics and Economic Education
  Research} \bibinfo{volume}{15} (\bibinfo{year}{2014}) \bibinfo{pages}{53}.
\bibitem[{Hong-ying et~al.(2008)Hong-ying, Yan-lin, Yong-juan
  et~al.}]{hong2008variations}
\bibinfo{author}{L.~Hong-ying}, \bibinfo{author}{H.~Yan-lin},
  \bibinfo{author}{Z.~Yong-juan}, et~al.,
\newblock \bibinfo{title}{Variations trend of grain yield per unit area and
  fertilizer application systems in liaoning province},
\newblock \bibinfo{journal}{System Sciences and Comprehensive Studies in
  Agriculture} \bibinfo{volume}{24} (\bibinfo{year}{2008})
  \bibinfo{pages}{408--410}.
\bibitem[{ZHANG and ZHANG(2007)}]{zhang2007application}
\bibinfo{author}{X.-j. ZHANG}, \bibinfo{author}{X.-l. ZHANG},
\newblock \bibinfo{title}{Application of time series analysis model on total
  corn yield of shandong province [j]},
\newblock \bibinfo{journal}{Research of Soil and Water Conservation}
  \bibinfo{volume}{3} (\bibinfo{year}{2007}) \bibinfo{pages}{098}.
\bibitem[{Yu and Babcock(2011)}]{yu2011estimating}
\bibinfo{author}{T.~Yu}, \bibinfo{author}{B.~A. Babcock},
\newblock \bibinfo{title}{Estimating non-linear weather impacts on corn
  yield—a bayesian approach}  (\bibinfo{year}{2011}).
\bibitem[{Chmielewski and Potts(1995)}]{chmielewski1995relationship}
\bibinfo{author}{F.-M. Chmielewski}, \bibinfo{author}{J.~M. Potts},
\newblock \bibinfo{title}{The relationship between crop yields from an
  experiment in southern england and long-term climate variations},
\newblock \bibinfo{journal}{Agricultural and Forest Meteorology}
  \bibinfo{volume}{73} (\bibinfo{year}{1995}) \bibinfo{pages}{43--66}.
\bibitem[{Jeong et~al.(2016)Jeong, Resop, Mueller, Fleisher, Yun, Butler,
  Timlin, Shim, Gerber, Reddy et~al.}]{jeong2016random}
\bibinfo{author}{J.~H. Jeong}, \bibinfo{author}{J.~P. Resop},
  \bibinfo{author}{N.~D. Mueller}, \bibinfo{author}{D.~H. Fleisher},
  \bibinfo{author}{K.~Yun}, \bibinfo{author}{E.~E. Butler},
  \bibinfo{author}{D.~J. Timlin}, \bibinfo{author}{K.-M. Shim},
  \bibinfo{author}{J.~S. Gerber}, \bibinfo{author}{V.~R. Reddy}, et~al.,
\newblock \bibinfo{title}{Random forests for global and regional crop yield
  predictions},
\newblock \bibinfo{journal}{PLoS One} \bibinfo{volume}{11}
  (\bibinfo{year}{2016}) \bibinfo{pages}{e0156571}.
\bibitem[{Everingham et~al.(2016)Everingham, Sexton, Skocaj, and
  Inman-Bamber}]{everingham2016accurate}
\bibinfo{author}{Y.~Everingham}, \bibinfo{author}{J.~Sexton},
  \bibinfo{author}{D.~Skocaj}, \bibinfo{author}{G.~Inman-Bamber},
\newblock \bibinfo{title}{Accurate prediction of sugarcane yield using a random
  forest algorithm},
\newblock \bibinfo{journal}{Agronomy for sustainable development}
  \bibinfo{volume}{36} (\bibinfo{year}{2016}) \bibinfo{pages}{27}.
\bibitem[{Dahikar and Rode(2014)}]{dahikar2014agricultural}
\bibinfo{author}{S.~S. Dahikar}, \bibinfo{author}{S.~V. Rode},
\newblock \bibinfo{title}{Agricultural crop yield prediction using artificial
  neural network approach},
\newblock \bibinfo{journal}{International Journal of Innovative Research in
  Electrical, Electronics, Instrumentation and Control Engineering}
  \bibinfo{volume}{2} (\bibinfo{year}{2014}) \bibinfo{pages}{683--686}.
\bibitem[{Fang et~al.(2018)Fang, Xu, Yang, and Qin}]{fang2018quantile}
\bibinfo{author}{Y.~Fang}, \bibinfo{author}{P.~Xu}, \bibinfo{author}{J.~Yang},
  \bibinfo{author}{Y.~Qin},
\newblock \bibinfo{title}{A quantile regression forest based method to predict
  drug response and assess prediction reliability},
\newblock \bibinfo{journal}{PloS one} \bibinfo{volume}{13}
  (\bibinfo{year}{2018}) \bibinfo{pages}{e0205155}.
\bibitem[{Jiang et~al.(2017)Jiang, Wu, and Peng}]{jiang2017semi}
\bibinfo{author}{F.~Jiang}, \bibinfo{author}{W.~Wu}, \bibinfo{author}{Z.~Peng},
\newblock \bibinfo{title}{A semi-parametric quantile regression random forest
  approach for evaluating muti-period value at risk},
\newblock in: \bibinfo{booktitle}{Control Conference (CCC), 2017 36th Chinese},
  \bibinfo{organization}{IEEE}, pp. \bibinfo{pages}{5642--5646}.
\bibitem[{Chatfield(2000)}]{chatfield2000time}
\bibinfo{author}{C.~Chatfield}, \bibinfo{title}{Time-series forecasting},
  \bibinfo{publisher}{Chapman and Hall/CRC}, \bibinfo{year}{2000}.
\bibitem[{Deser et~al.(2012)Deser, Phillips, Bourdette, and
  Teng}]{deser2012uncertainty}
\bibinfo{author}{C.~Deser}, \bibinfo{author}{A.~Phillips},
  \bibinfo{author}{V.~Bourdette}, \bibinfo{author}{H.~Teng},
\newblock \bibinfo{title}{Uncertainty in climate change projections: the role
  of internal variability},
\newblock \bibinfo{journal}{Climate dynamics} \bibinfo{volume}{38}
  (\bibinfo{year}{2012}) \bibinfo{pages}{527--546}.
\bibitem[{Reilly et~al.(2001)Reilly, Stone, Forest, Webster, Jacoby, and
  Prinn}]{reilly2001uncertainty}
\bibinfo{author}{J.~Reilly}, \bibinfo{author}{P.~H. Stone},
  \bibinfo{author}{C.~E. Forest}, \bibinfo{author}{M.~D. Webster},
  \bibinfo{author}{H.~D. Jacoby}, \bibinfo{author}{R.~G. Prinn},
  \bibinfo{title}{Uncertainty and climate change assessments},
  \bibinfo{year}{2001}.
\bibitem[{Wand and Jones(1994)}]{wand1994kernel}
\bibinfo{author}{M.~P. Wand}, \bibinfo{author}{M.~C. Jones},
  \bibinfo{title}{Kernel smoothing}, \bibinfo{publisher}{Chapman and Hall/CRC},
  \bibinfo{year}{1994}.
\bibitem[{Barnwal and Kotani(2013)}]{barnwal2013climatic}
\bibinfo{author}{P.~Barnwal}, \bibinfo{author}{K.~Kotani},
\newblock \bibinfo{title}{Climatic impacts across agricultural crop yield
  distributions: An application of quantile regression on rice crops in andhra
  pradesh, india},
\newblock \bibinfo{journal}{Ecological Economics} \bibinfo{volume}{87}
  (\bibinfo{year}{2013}) \bibinfo{pages}{95--109}.
\bibitem[{Wang(2012)}]{wang2012bayesian}
\bibinfo{author}{J.~Wang},
\newblock \bibinfo{title}{Bayesian quantile regression for parametric nonlinear
  mixed effects models},
\newblock \bibinfo{journal}{Statistical Methods \& Applications}
  \bibinfo{volume}{21} (\bibinfo{year}{2012}) \bibinfo{pages}{279--295}.
\bibitem[{Meinshausen(2006)}]{meinshausen2006quantile}
\bibinfo{author}{N.~Meinshausen},
\newblock \bibinfo{title}{Quantile regression forests},
\newblock \bibinfo{journal}{Journal of Machine Learning Research}
  \bibinfo{volume}{7} (\bibinfo{year}{2006}) \bibinfo{pages}{983--999}.
\bibitem[{Jones et~al.(1996)Jones, Marron, and Sheather}]{jones1996brief}
\bibinfo{author}{M.~C. Jones}, \bibinfo{author}{J.~S. Marron},
  \bibinfo{author}{S.~J. Sheather},
\newblock \bibinfo{title}{A brief survey of bandwidth selection for density
  estimation},
\newblock \bibinfo{journal}{Journal of the American statistical association}
  \bibinfo{volume}{91} (\bibinfo{year}{1996}) \bibinfo{pages}{401--407}.
\bibitem[{Koenker and Bassett(1978)}]{koenker1978regression}
\bibinfo{author}{R.~Koenker}, \bibinfo{author}{G.~Bassett},
\newblock \bibinfo{title}{Regression quantiles. econometrica 46 33--50},
\newblock \bibinfo{journal}{Mathematical Reviews (MathSciNet): MR474644 Digital
  Object Identifier: doi} \bibinfo{volume}{10} (\bibinfo{year}{1978})
  \bibinfo{pages}{1913643}.
\bibitem[{Lin and Jeon(2006)}]{lin2006random}
\bibinfo{author}{Y.~Lin}, \bibinfo{author}{Y.~Jeon},
\newblock \bibinfo{title}{Random forests and adaptive nearest neighbors},
\newblock \bibinfo{journal}{Journal of the American Statistical Association}
  \bibinfo{volume}{101} (\bibinfo{year}{2006}) \bibinfo{pages}{578--590}.
\bibitem[{Epanechnikov(1969)}]{epanechnikov1969non}
\bibinfo{author}{V.~A. Epanechnikov},
\newblock \bibinfo{title}{Non-parametric estimation of a multivariate
  probability density},
\newblock \bibinfo{journal}{Theory of Probability \& Its Applications}
  \bibinfo{volume}{14} (\bibinfo{year}{1969}) \bibinfo{pages}{153--158}.
\bibitem[{Sheather and Jones(1991)}]{sheather1991reliable}
\bibinfo{author}{S.~J. Sheather}, \bibinfo{author}{M.~C. Jones},
\newblock \bibinfo{title}{A reliable data-based bandwidth selection method for
  kernel density estimation},
\newblock \bibinfo{journal}{Journal of the Royal Statistical Society: Series B
  (Methodological)} \bibinfo{volume}{53} (\bibinfo{year}{1991})
  \bibinfo{pages}{683--690}.
\bibitem[{FAO(2010)}]{fao1}
\bibinfo{author}{FAO}, \bibinfo{title}{Crop calendar - an information tool for
  seed security},
  \bibinfo{howpublished}{$http://www.fao.org/agriculture/seed/cropcalendar/welcome.do$},
  \bibinfo{year}{2010}. \bibinfo{note}{(accessed 26 September 2018)}.
\bibitem[{Alexandrov and Hoogenboom(2000)}]{alexandrov2000impact}
\bibinfo{author}{V.~Alexandrov}, \bibinfo{author}{G.~Hoogenboom},
\newblock \bibinfo{title}{The impact of climate variability and change on crop
  yield in bulgaria},
\newblock \bibinfo{journal}{Agricultural and forest meteorology}
  \bibinfo{volume}{104} (\bibinfo{year}{2000}) \bibinfo{pages}{315--327}.

\end{thebibliography}


\begin{thebibliography}{25}
\expandafter\ifx\csname natexlab\endcsname\relax\def\natexlab#1{#1}\fi
\providecommand{\bibinfo}[2]{#2}
\ifx\xfnm\relax \def\xfnm[#1]{\unskip,\space#1}\fi
\bibitem[{Rana and Rana(2014)}]{rana2014}
\bibinfo{author}{S.~S. Rana}, \bibinfo{author}{R.~S. Rana},
  \bibinfo{title}{ADVANCES IN CROP GROWTH AND PRODUCTIVITY},
  \bibinfo{type}{Technical Report}, Publication of the Department of Agronomy,
  CSK Himachal Pradesh Krishi Vishvavidyalaya, Palampur, India,
  \bibinfo{year}{2014}.
\bibitem[{Gyamerah et~al.(2018)Gyamerah, Ngare, and Ikpe}]{gyamerah2018regime}
\bibinfo{author}{S.~A. Gyamerah}, \bibinfo{author}{P.~Ngare},
  \bibinfo{author}{D.~Ikpe},
\newblock \bibinfo{title}{Regime-switching temperature dynamics model for
  weather derivatives},
\newblock \bibinfo{journal}{International Journal of Stochastic Analysis}
  \bibinfo{volume}{2018} (\bibinfo{year}{2018}).
\bibitem[{Delinc{\'e} et~al.(2017)}]{delince2017recent}
\bibinfo{author}{J.~Delinc{\'e}}, et~al.,
\newblock \bibinfo{title}{Recent practices and advances for amis crop yield
  forecasting at farm and parcel level: a review.},
\newblock \bibinfo{journal}{Recent practices and advances for AMIS crop yield
  forecasting at farm and parcel level: a review.}  (\bibinfo{year}{2017}).
\bibitem[{Shi et~al.(2013)Shi, Tao, and Zhang}]{shi2013review}
\bibinfo{author}{W.~Shi}, \bibinfo{author}{F.~Tao}, \bibinfo{author}{Z.~Zhang},
\newblock \bibinfo{title}{A review on statistical models for identifying
  climate contributions to crop yields},
\newblock \bibinfo{journal}{Journal of geographical sciences}
  \bibinfo{volume}{23} (\bibinfo{year}{2013}) \bibinfo{pages}{567--576}.
\bibitem[{Michler et~al.(2015)Michler, Viens, Shively et~al.}]{michler2015risk}
\bibinfo{author}{J.~D. Michler}, \bibinfo{author}{F.~G. Viens},
  \bibinfo{author}{G.~E. Shively}, et~al.,
\newblock \bibinfo{title}{Risk, agricultural production, and weather index
  insurance in village south asia},
\newblock in: \bibinfo{booktitle}{International Association of Agricultural
  Economists 2015 International Conference of Agricultural Economists, Milan,
  Italy, August}, pp. \bibinfo{pages}{8--14}.
\bibitem[{Lobell and Burke(2010)}]{lobell2010use}
\bibinfo{author}{D.~B. Lobell}, \bibinfo{author}{M.~B. Burke},
\newblock \bibinfo{title}{On the use of statistical models to predict crop
  yield responses to climate change},
\newblock \bibinfo{journal}{Agricultural and Forest Meteorology}
  \bibinfo{volume}{150} (\bibinfo{year}{2010}) \bibinfo{pages}{1443--1452}.
\bibitem[{Choudhury and Jones(2014)}]{choudhury2014crop}
\bibinfo{author}{A.~Choudhury}, \bibinfo{author}{J.~Jones},
\newblock \bibinfo{title}{Crop yield prediction using time series models},
\newblock \bibinfo{journal}{Journal of Economics and Economic Education
  Research} \bibinfo{volume}{15} (\bibinfo{year}{2014}) \bibinfo{pages}{53}.
\bibitem[{Hong-ying et~al.(2008)Hong-ying, Yan-lin, Yong-juan
  et~al.}]{hong2008variations}
\bibinfo{author}{L.~Hong-ying}, \bibinfo{author}{H.~Yan-lin},
  \bibinfo{author}{Z.~Yong-juan}, et~al.,
\newblock \bibinfo{title}{Variations trend of grain yield per unit area and
  fertilizer application systems in liaoning province},
\newblock \bibinfo{journal}{System Sciences and Comprehensive Studies in
  Agriculture} \bibinfo{volume}{24} (\bibinfo{year}{2008})
  \bibinfo{pages}{408--410}.
\bibitem[{ZHANG and ZHANG(2007)}]{zhang2007application}
\bibinfo{author}{X.-j. ZHANG}, \bibinfo{author}{X.-l. ZHANG},
\newblock \bibinfo{title}{Application of time series analysis model on total
  corn yield of shandong province [j]},
\newblock \bibinfo{journal}{Research of Soil and Water Conservation}
  \bibinfo{volume}{3} (\bibinfo{year}{2007}) \bibinfo{pages}{098}.
\bibitem[{Jeong et~al.(2016)Jeong, Resop, Mueller, Fleisher, Yun, Butler,
  Timlin, Shim, Gerber, Reddy et~al.}]{jeong2016random}
\bibinfo{author}{J.~H. Jeong}, \bibinfo{author}{J.~P. Resop},
  \bibinfo{author}{N.~D. Mueller}, \bibinfo{author}{D.~H. Fleisher},
  \bibinfo{author}{K.~Yun}, \bibinfo{author}{E.~E. Butler},
  \bibinfo{author}{D.~J. Timlin}, \bibinfo{author}{K.-M. Shim},
  \bibinfo{author}{J.~S. Gerber}, \bibinfo{author}{V.~R. Reddy}, et~al.,
\newblock \bibinfo{title}{Random forests for global and regional crop yield
  predictions},
\newblock \bibinfo{journal}{PLoS One} \bibinfo{volume}{11}
  (\bibinfo{year}{2016}) \bibinfo{pages}{e0156571}.
\bibitem[{Everingham et~al.(2016)Everingham, Sexton, Skocaj, and
  Inman-Bamber}]{everingham2016accurate}
\bibinfo{author}{Y.~Everingham}, \bibinfo{author}{J.~Sexton},
  \bibinfo{author}{D.~Skocaj}, \bibinfo{author}{G.~Inman-Bamber},
\newblock \bibinfo{title}{Accurate prediction of sugarcane yield using a random
  forest algorithm},
\newblock \bibinfo{journal}{Agronomy for sustainable development}
  \bibinfo{volume}{36} (\bibinfo{year}{2016}) \bibinfo{pages}{27}.
\bibitem[{Dahikar and Rode(2014)}]{dahikar2014agricultural}
\bibinfo{author}{S.~S. Dahikar}, \bibinfo{author}{S.~V. Rode},
\newblock \bibinfo{title}{Agricultural crop yield prediction using artificial
  neural network approach},
\newblock \bibinfo{journal}{International Journal of Innovative Research in
  Electrical, Electronics, Instrumentation and Control Engineering}
  \bibinfo{volume}{2} (\bibinfo{year}{2014}) \bibinfo{pages}{683--686}.
\bibitem[{Fang et~al.(2018)Fang, Xu, Yang, and Qin}]{fang2018quantile}
\bibinfo{author}{Y.~Fang}, \bibinfo{author}{P.~Xu}, \bibinfo{author}{J.~Yang},
  \bibinfo{author}{Y.~Qin},
\newblock \bibinfo{title}{A quantile regression forest based method to predict
  drug response and assess prediction reliability},
\newblock \bibinfo{journal}{PloS one} \bibinfo{volume}{13}
  (\bibinfo{year}{2018}) \bibinfo{pages}{e0205155}.
\bibitem[{Jiang et~al.(2017)Jiang, Wu, and Peng}]{jiang2017semi}
\bibinfo{author}{F.~Jiang}, \bibinfo{author}{W.~Wu}, \bibinfo{author}{Z.~Peng},
\newblock \bibinfo{title}{A semi-parametric quantile regression random forest
  approach for evaluating muti-period value at risk},
\newblock in: \bibinfo{booktitle}{Control Conference (CCC), 2017 36th Chinese},
  \bibinfo{organization}{IEEE}, pp. \bibinfo{pages}{5642--5646}.
\bibitem[{Deser et~al.(2012)Deser, Phillips, Bourdette, and
  Teng}]{deser2012uncertainty}
\bibinfo{author}{C.~Deser}, \bibinfo{author}{A.~Phillips},
  \bibinfo{author}{V.~Bourdette}, \bibinfo{author}{H.~Teng},
\newblock \bibinfo{title}{Uncertainty in climate change projections: the role
  of internal variability},
\newblock \bibinfo{journal}{Climate dynamics} \bibinfo{volume}{38}
  (\bibinfo{year}{2012}) \bibinfo{pages}{527--546}.
\bibitem[{Reilly et~al.(2001)Reilly, Stone, Forest, Webster, Jacoby, and
  Prinn}]{reilly2001uncertainty}
\bibinfo{author}{J.~Reilly}, \bibinfo{author}{P.~H. Stone},
  \bibinfo{author}{C.~E. Forest}, \bibinfo{author}{M.~D. Webster},
  \bibinfo{author}{H.~D. Jacoby}, \bibinfo{author}{R.~G. Prinn},
  \bibinfo{title}{Uncertainty and climate change assessments},
  \bibinfo{year}{2001}.
\bibitem[{Barnwal and Kotani(2013)}]{barnwal2013climatic}
\bibinfo{author}{P.~Barnwal}, \bibinfo{author}{K.~Kotani},
\newblock \bibinfo{title}{Climatic impacts across agricultural crop yield
  distributions: An application of quantile regression on rice crops in andhra
  pradesh, india},
\newblock \bibinfo{journal}{Ecological Economics} \bibinfo{volume}{87}
  (\bibinfo{year}{2013}) \bibinfo{pages}{95--109}.
\bibitem[{Wang(2012)}]{wang2012bayesian}
\bibinfo{author}{J.~Wang},
\newblock \bibinfo{title}{Bayesian quantile regression for parametric nonlinear
  mixed effects models},
\newblock \bibinfo{journal}{Statistical Methods \& Applications}
  \bibinfo{volume}{21} (\bibinfo{year}{2012}) \bibinfo{pages}{279--295}.
\bibitem[{Meinshausen(2006)}]{meinshausen2006quantile}
\bibinfo{author}{N.~Meinshausen},
\newblock \bibinfo{title}{Quantile regression forests},
\newblock \bibinfo{journal}{Journal of Machine Learning Research}
  \bibinfo{volume}{7} (\bibinfo{year}{2006}) \bibinfo{pages}{983--999}.
\bibitem[{Koenker and Bassett(1978)}]{koenker1978regression}
\bibinfo{author}{R.~Koenker}, \bibinfo{author}{G.~Bassett},
\newblock \bibinfo{title}{Regression quantiles. econometrica 46 33--50},
\newblock \bibinfo{journal}{Mathematical Reviews (MathSciNet): MR474644 Digital
  Object Identifier: doi} \bibinfo{volume}{10} (\bibinfo{year}{1978})
  \bibinfo{pages}{1913643}.
\bibitem[{Lin and Jeon(2006)}]{lin2006random}
\bibinfo{author}{Y.~Lin}, \bibinfo{author}{Y.~Jeon},
\newblock \bibinfo{title}{Random forests and adaptive nearest neighbors},
\newblock \bibinfo{journal}{Journal of the American Statistical Association}
  \bibinfo{volume}{101} (\bibinfo{year}{2006}) \bibinfo{pages}{578--590}.
\bibitem[{Epanechnikov(1969)}]{epanechnikov1969non}
\bibinfo{author}{V.~A. Epanechnikov},
\newblock \bibinfo{title}{Non-parametric estimation of a multivariate
  probability density},
\newblock \bibinfo{journal}{Theory of Probability \& Its Applications}
  \bibinfo{volume}{14} (\bibinfo{year}{1969}) \bibinfo{pages}{153--158}.
\bibitem[{Wand and Jones(1994)}]{wand1994kernel}
\bibinfo{author}{M.~P. Wand}, \bibinfo{author}{M.~C. Jones},
  \bibinfo{title}{Kernel smoothing}, \bibinfo{publisher}{Chapman and Hall/CRC},
  \bibinfo{year}{1994}.
\bibitem[{Jones et~al.(1996)Jones, Marron, and Sheather}]{jones1996brief}
\bibinfo{author}{M.~C. Jones}, \bibinfo{author}{J.~S. Marron},
  \bibinfo{author}{S.~J. Sheather},
\newblock \bibinfo{title}{A brief survey of bandwidth selection for density
  estimation},
\newblock \bibinfo{journal}{Journal of the American statistical association}
  \bibinfo{volume}{91} (\bibinfo{year}{1996}) \bibinfo{pages}{401--407}.
\bibitem[{Sheather and Jones(1991)}]{sheather1991reliable}
\bibinfo{author}{S.~J. Sheather}, \bibinfo{author}{M.~C. Jones},
\newblock \bibinfo{title}{A reliable data-based bandwidth selection method for
  kernel density estimation},
\newblock \bibinfo{journal}{Journal of the Royal Statistical Society: Series B
  (Methodological)} \bibinfo{volume}{53} (\bibinfo{year}{1991})
  \bibinfo{pages}{683--690}.

\end{thebibliography}

\end{document}